\documentclass[8.5pt,twoside,twocolumn]{article}
\usepackage[super,sort&compress,comma]{natbib}
\usepackage{mhchem}
\usepackage{times,mathptmx}
\usepackage{sectsty}
\usepackage{color}
\usepackage{balance}
\usepackage{bm}
\usepackage{epsfig}
\usepackage{graphicx} 
\usepackage{lastpage}
\usepackage[format=plain,justification=raggedright,singlelinecheck=false,font=small,labelfont=bf,labelsep=space]{caption}
\usepackage{fancyhdr}
\usepackage[T1]{fontenc}
\begin{document}

\thispagestyle{plain}
\fancypagestyle{plain}{
\fancyhead[L]{\includegraphics[height=8pt]{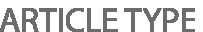}}
\fancyhead[C]{\hspace{-1cm}\includegraphics[height=20pt]{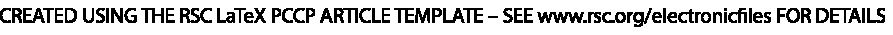}}
\fancyhead[R]{\includegraphics[height=10pt]{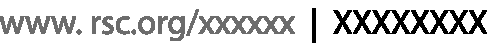}\vspace{-0.2cm}}
\renewcommand{\headrulewidth}{1pt}}
\renewcommand{\thefootnote}{\fnsymbol{footnote}}
\renewcommand\footnoterule{\vspace*{1pt}%
\hrule width 3.4in height 0.4pt \vspace*{5pt}}
\setcounter{secnumdepth}{5}

\makeatletter
\def\subsubsection{\@startsection{subsubsection}{3}{10pt}{-1.25ex plus -1ex minus -.1ex}{0ex plus 0ex}{\normalsize\bf}}
\def\paragraph{\@startsection{paragraph}{4}{10pt}{-1.25ex plus -1ex minus -.1ex}{0ex plus 0ex}{\normalsize\textit}}
\renewcommand\@biblabel[1]{#1}
\renewcommand\@makefntext[1]%
{\noindent\makebox[0pt][r]{\@thefnmark\,}#1}
\makeatother
\renewcommand{\figurename}{\small{Fig.}~}
\sectionfont{\large}
\subsectionfont{\normalsize}

\fancyfoot{}
\fancyfoot[LO,RE]{\vspace{-7pt}\includegraphics[height=9pt]{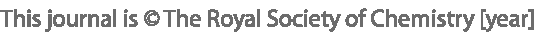}}
\fancyfoot[CO]{\vspace{-7.2pt}\hspace{12.2cm}\includegraphics{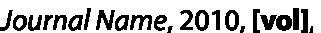}}
\fancyfoot[CE]{\vspace{-7.5pt}\hspace{-13.5cm}\includegraphics{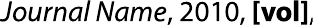}}
\fancyfoot[RO]{\footnotesize{\sffamily{1--\pageref{LastPage} ~\textbar  \hspace{2pt}\thepage}}}
\fancyfoot[LE]{\footnotesize{\sffamily{\thepage~\textbar\hspace{3.45cm} 1--\pageref{LastPage}}}}
\fancyhead{}
\renewcommand{\headrulewidth}{1pt}
\renewcommand{\footrulewidth}{1pt}
\setlength{\arrayrulewidth}{1pt}
\setlength{\columnsep}{6.5mm}
\setlength\bibsep{1pt}

\twocolumn[
  \begin{@twocolumnfalse}
\noindent\LARGE{\textbf{Discrete Boltzmann modeling of multiphase flows: hydrodynamic and thermodynamic non-equilibrium effects}}
\vspace{0.6cm}

\noindent\large{\textbf{Yanbiao Gan,\textit{$^{a,b}$} Aiguo Xu,$^{\ast}$\textit{$^{a,c,d}$}
Guangcai Zhang\textit{$^{a,d}$} and
Sauro Succi\textit{$^{e}$}}}\vspace{0.5cm}

\noindent\textit{\small{\textbf{Received Xth XXXXXXXXXX 20XX, \newline
}}}

\noindent \textbf{\small{DOI: }}
\vspace{0.6cm}

\noindent \normalsize{
A discrete Boltzmann model (DBM) is developed to investigate the hydrodynamic and thermodynamic non-equilibrium (TNE) effects in phase separation processes. The interparticle force drives changes and the gradient force, induced by gradients of macroscopic quantities, opposes them. In this paper, we investigate the interplay between them by providing detailed inspection of various non-equilibrium observables. Based on the TNE features, we define a TNE strength which roughly estimates the deviation amplitude from the thermodynamic equilibrium.
The time evolution of the TNE intensity provides a convenient and efficient physical criterion to discriminate the stages of the spinodal decomposition and domain growth. Via the DBM simulation and this criterion, we quantitatively study the effects of latent heat and surface tension on phase separation. It is found that, the TNE strength attains its maximum at the end of the spinodal decomposition stage, and it decreases when the latent heat increases from zero.
The surface tension effects are threefold, to prolong the duration of the spinodal decomposition stage, decrease the maximum TNE intensity, and accelerate the speed of the domain growth stage.}
\vspace{0.5cm}
 \end{@twocolumnfalse}
  ]

\section{Introduction}


\footnotetext{\textit{$^{a}$~National Key Laboratory of Computational Physics,
Institute of Applied Physics and Computational Mathematics, P. O.
Box 8009-26, Beijing, PRC. E-mail:Xu\_Aiguo@iapcm.ac.cn}}
\footnotetext{\textit{$^{b}$~North China Institute of Aerospace Engineering, Langfang, PRC. }}
\footnotetext{\textit{$^{c}$~Center for Applied Physics and Technology, MOE Key Center for High Energy
Density Physics Simulations, College of Engineering, Peking University,
Beijing, PRC. }}
\footnotetext{\textit{$^{d}$~State Key Laboratory of Theoretical Physics, Institute of Theoretical Physics, Chinese Academy of Sciences, Beijing, PRC. }}
\footnotetext{\textit{$^{e}$~Istituto Applicazioni Calcolo, Via dei Taurini, Roma, Italy.}}



Owing to the existence of complex interparticle interactions at the
microscopic level and nonlinear interfaces between various phases/components
at the macroscopic level, the hydrodynamic non-equilibrium (HNE) and thermodynamic
non-equilibrium (TNE) effects play a major role in shaping up the essential features of dynamic relaxation phenomena in multiphase flow systems. The HNE and TNE show the features of the system in different aspects. The traditional Navier-Stokes model describes well weak HNE, but encounters difficulties in describing the TNE. To this purpose, a model based on the Boltzmann equation is preferable.

In the past two decades, as a special discretization of the Boltzmann equation, the lattice Boltzmann method has carried substantially forward on the physical understanding of multiphase flows.\cite{Succi,Succi-group,Yeomans-group,SC-group,Cates-group,
Sega,Basagaoglu,
Gonnella-group,EPJB-Sofonea,Xu-PRE2011-EPL2012,
JCP-2005,Yuan-POF,Liqing-PRE2013,Fang-SR} In recent studies,\cite{our-group,Succi-JCP-2015,Xu-PRE2015} the lattice Boltzmann method was developed to probe the trans- and supercritical fluid behaviors or both the HNE and TNE simultaneously in complex flows, which bring some new physical insights into the fine structures in the system. Such an extended lattice Boltzmann kinetic model or discrete Boltzmann model (DBM) should follow more strictly some necessary kinetic moment relations of the equilibrium distribution function $f_{ki}^{eq}$. Thus, besides recovering the Navier-Stokes equation, it describes also the evolution of some non-conserved physical quantities, for example, the difference of the energies in various degrees of freedom, the energy flux, etc, in the continuum limit. The TNE can be simply measured by the differences between the non-conserved kinetic moments of $f_{ki}$ and $f_{ki}^{eq}$, where $f_{ki}$ is the discrete distribution function of the $k$-th group of particles. In other words, the TNE behaviors can be extracted dynamically from the DBM simulation without employing additional analysis tools.

In this paper, we present a DBM for multiphase flows with flexible density ratio,
formulate new quantitative measures of TNE effects in the system, and utilize them to probe the phase separation process.

\section{DBM for thermal multiphase flows with flexible density ratio}

In 2007, Gonnella, Lamura and Sofonea (GLS) proposed a thermal lattice Boltzmann
model for multiphase flows through introducing an appropriate interparticle force  into the lattice Boltzmann equation.\cite{Gonnella-group} In 2011, some of the present authors developed the GLS model by implementing the fast Fourier transform and its inverse to calculate the spatial derivatives.\cite{Xu-PRE2011-EPL2012}
As a result, the total energy conservation can be better held and the spurious velocities are refrained to a negligible scale in real simulations.
In this work we further improve the DBM in two sides, insert a more practical equation of state and supplement  a methodology to investigate the out-of-equilibrium features in the multiphase flow.

The GLS-lattice Boltzmann equation reads as follows:
\begin{equation}
\frac{\partial f_{ki}}{\partial t}+\mathbf{v}_{ki} \cdot \frac{\partial f_{ki}
}{\partial \mathbf{r}}=-\frac{1}{\tau} [f_{ki}-f_{ki}^{eq}]
+I_{ki}\text{,}  \label{GLS-LB}
\end{equation}%
where $I_{ki}$ takes the following form:
\begin{equation}
I_{ki}=-[A+ \mathbf{B} \cdot (\mathbf{v}_{ki}-\mathbf{u})+(C+C_{q})(\mathbf{v}_{ki}
-\mathbf{u})^{2}]f_{ki}^{eq}\text{,}  \label{iki}
\end{equation}%
with \begin{equation}
A=-2(C+C_{q})T\text{,}  \label{AAA}
\end{equation}%
\begin{equation}
\mathbf{B}=\frac{1}{\rho T}[\mathbf{\nabla}(P^{\text{vdw}}-\rho T)+\mathbf{\nabla}\cdot
\boldsymbol{\Lambda }-\mathbf{\nabla}(\zeta \mathbf{\nabla}\cdot \mathbf{u})],
\label{BBB}
\end{equation}%
\begin{eqnarray}
C &=&\frac{1}{2\rho T^{2}}\{(P^{\text{vdw}}-\rho T)\mathbf{\nabla}\cdot \mathbf{u}+%
\boldsymbol{\Lambda \colon \mathbf{\nabla}u}-\zeta (\mathbf{\nabla}\cdot \mathbf{u})^{2}
\notag \\
&&+\frac{9}{8}\rho ^{2}\mathbf{\nabla}\cdot \mathbf{u}+K[-\frac{1}{2}(
\mathbf{\nabla}\rho \cdot \mathbf{\nabla}\rho )\mathbf{\nabla}\cdot \mathbf{u}  \notag \\
&&-\rho \mathbf{\nabla}\rho \cdot \mathbf{\nabla}(\mathbf{\nabla}\cdot \mathbf{u
})-\mathbf{\nabla}\rho \mathbf{\cdot \mathbf{\nabla}u\cdot }\mathbf{\nabla}\rho
]\}\text{,}  \label{CC}
\end{eqnarray}%
\begin{equation}
C_{q}=\frac{1}{2\rho T^{2}}\mathbf{\nabla}\cdot \lbrack 2q\rho T\mathbf{\nabla}T]%
\text{.}  \label{CCQQ}
\end{equation}%
Here $\rho $, $\mathbf{u}$, $T$ are the local density, velocity,
temperature, respectively.
$\boldsymbol{\Lambda }=K\mathbf{\nabla}\rho \mathbf{\nabla}
\rho -K(\rho \nabla ^{2}\rho +\left\vert \mathbf{\nabla}\rho \right\vert ^{2}/2)%
\mathbf{I}-[\rho T\mathbf{\nabla}\rho \cdot \mathbf{\nabla}(K/T)]\mathbf{I}$ is the
contribution of density gradient to pressure tensor, $\mathbf{I}$ is the
unit tensor, $K$ is the surface tension coefficient. $\zeta$ is the bulk viscosity.
In the continuum limit, GLS model corresponds to, whereas is beyond, the thermohydrodynamic equations proposed by Onuki.\cite{Onuki-PRL}

It is noteworthy that GLS model utilizes the van der Waals equation of state: $P^{\text{vdw}}$ $=
\frac{3\rho T}{3-\rho }-\frac{9}{8}\rho ^{2}$ with fixed parameters. Due to numerical instabilities, the density ratio $R$ between the liquid and vapor phases
that the model can support is less than $10$. However, in practical
engineering applications and natural situations, $R$ can vary greatly. For
example, the density ratio of a liquid alloy system is close to $1$, but
that of water to steam is about $858$ under the standard conditions.
To improve the lattice Boltzmann model for simulating multiphase flows with large density ratio, extensive efforts have been made.\cite{JCP-2005,Yuan-POF,Liqing-PRE2013} Among them, Yuan and Schaefer's approach\cite{Yuan-POF} is straightforward and effective. The core idea is that, through rearranging the effective mass in the pseudopotential model, more realistic equations of state, such as Redlich-Kwong,\cite{RK} Peng-Robinson,\cite{PR} and Carnahan-Starling\cite{CS} equation of state could be incorporated into the lattice Boltzmann model, which dramatically increased the density ratio, decreased the spurious currents, thereby minimizing the thermodynamic inconsistency. Similarly, in this work, through modifying the forcing term, i.e.,
replacing the term $\frac{9}{8}\rho ^{2}\mathbf{\nabla}
\cdot \mathbf{u}$ in eqn (5) by $a\rho ^{2}\mathbf{\nabla}\cdot \mathbf{u}$,
the following Carnahan-Starling equation of state can be adopted
\begin{equation}
p^{\text{cs}}=\rho T\frac{1+\eta +\eta ^{2}-\eta ^{3}}{(1-\eta )^{3}}-a\rho ^{2},  \label{PCS}
\end{equation}%
with $\eta =b\rho /4$, $a$ and $b$ are the attraction and repulsion parameters. Subsequently, the total energy density becomes $e_{T}=\rho T-a\rho^{2}+K\left\vert \mathbf{\nabla}\rho \right\vert ^{2}/2+\rho u^{2}/2$.
The Carnahan-Starling equation of state modified the repulsive term of van der Waals equation of state and obtained a
more accurate representation for hard sphere interactions. The incorporation of this equation into the DBM belongs to an improvement of physical modeling. Chapman-Enskog analysis and the following numerical tests demonstrate that the revised model is thermodynamic consistent. Compared to models listed in refs.10-12, our model is a thermal and compressible one that can be used to probe both the HNE and TNE effects.

\section{Two kinds of non-equilibrium effects}

To recover the thermohydrodynamic equations at the Navier-Stokes level, GLS model uses the following seven kinetic moments,
\begin{equation}
\mathbf{M}_{0}^{\text{eq}}=\sum_{ki}f_{ki}^{eq}=\rho \text{,}
\label{moment1}
\end{equation}
\begin{equation}
\mathbf{M}_{1}^{\text{eq}}=\sum_{ki}f_{ki}^{eq}\mathbf{v}
_{ki}=\rho \mathbf{u}\text{,}  \label{moment2}
\end{equation}
\begin{equation}
\mathbf{M}_{2,0}^{\text{eq}}=\sum_{ki}\frac{1}{2}f_{ki}^{eq}\mathbf{v}%
_{ki}\cdot \mathbf{v}_{ki}=\rho (T+\frac{1}{2}\mathbf{u}\cdot \mathbf{u})%
\text{,}
\end{equation}
\begin{equation}
\mathbf{M}_{2}^{\text{eq}}=\sum_{ki}f_{ki}^{eq}\mathbf{v}_{ki}\mathbf{v}%
_{ki}=\rho (T\mathbf{I}+\mathbf{uu})\text{,}  \label{moment4}
\end{equation}
\begin{eqnarray}
\mathbf{M}_{3}^{\text{eq}}=\sum_{ki}f_{ki}^{eq}\mathbf{v}_{ki}\mathbf{v}%
_{ki}\mathbf{v}_{ki}
=\rho \lbrack T(\mathbf{u}_{\alpha }\mathbf{e}_{\beta }\mathbf{e}_{\gamma
}\delta _{\beta \gamma }+\mathbf{e}_{\alpha }\mathbf{u}_{\beta }\mathbf{e}%
_{\gamma }\delta_{\alpha \gamma} \notag \\
+\mathbf{e}_{\alpha }\mathbf{e}_{\beta }
\mathbf{u}_{\gamma }\delta_{\alpha \beta })+\mathbf{uuu}]\text{,}
\label{moment5}
\end{eqnarray}
\begin{equation}
\mathbf{M}_{3,1}^{\text{eq}}=\sum_{ki}\frac{1}{2}f_{ki}^{eq}\mathbf{v}%
_{ki}\cdot \mathbf{v}_{ki}\mathbf{v}_{ki}=\rho \mathbf{u}(2T+\frac{1}{2}%
\mathbf{u}\cdot \mathbf{u})\text{,}  \label{moment6}
\end{equation}%
\begin{eqnarray}
\mathbf{M}_{4,2}^{\text{eq}} =\sum_{ki}\frac{1}{2}f_{ki}^{eq}\mathbf{v}%
_{ki}\cdot \mathbf{v}_{ki}\mathbf{v}_{ki}\mathbf{v}_{ki}=\rho \lbrack (2T+%
\frac{\mathbf{u\cdot u}}{2})T\mathbf{I}  \notag \\
+\mathbf{uu}(3T+\frac{\mathbf{u\cdot u}}{2})] \text{,} \label{moment7}
\end{eqnarray}%
where $\mathbf{M}_{m,n}^{\text{eq}}$ stands for that the $m$-th order tensor is contracted to a $n$-th order one.
Among the seven kinetic moment relations, only for the first three ones, $f_{ki}^{eq}$ can be replaced by $f_{ki}$, which means that in or out of the equilibrium, the mass, momentum and energy conservations are kept. Replacing $f_{ki}^{eq}$ by $f_{ki}$ in eqns \eqref{moment4}-\eqref{moment7} will lead to the imbalance and the deviation
\begin{equation}
\boldsymbol{\Delta}_{n} =\mathbf{M}_{n}(f_{ki})-\mathbf{M}_{n}^{\text{eq}}(f_{ki}^{eq}),
\end{equation}
which can be used to measures the departure of the system from the local thermodynamic equilibrium.

For an ideal gas system, the HNE and TNE effects are only induced by gradients of macroscopic quantities, also referred to gradient force. For multiphase flow system, the existence of interparticle force makes the situation a little more complex. The force term in the DBM equation works as the second driving force. Especially, the right-hand side of eqn (\ref{GLS-LB}) can be reorganized as
\begin{equation}
\text{RHS}=-\frac{1}{\tau }[f_{ki}-(1+\tau \theta
)f_{ki}^{eq}]=-\frac{1}{\tau }[f_{ki}-f_{ki}^{eq,\text{NEW}}]\text{,}
\end{equation}%
where $\theta=-[A+ \mathbf{B} \cdot (\mathbf{v}_{ki}-\mathbf{u})+(C+C_{q})(\mathbf{v}_{ki}
-\mathbf{u})^{2}]$, $f_{ki}^{eq,\text{NEW}}=(1+\tau \theta)f_{ki}^{eq}$
can be considered as a new equilibrium state shifted by the interparticle force. Thus,
\begin{equation}
\boldsymbol{\Delta }_{n}^{F}=\mathbf{M}_{n}(\tau \theta f_{ki}^{eq})=%
\mathbf{M}_{n}(\tau I_{ki})
\end{equation}%
are the non-equilibrium effects induced by the interparticle force,
and what we measured from $f_{ki}$ and $f^{eq}_{ki}$,
\begin{equation}
\boldsymbol{\Delta}_{n}=\mathbf{M}_{n}(f_{ki})-\mathbf{M}_{n}^{\text{eq}}(f^{eq}_{ki})=%
\boldsymbol{\Delta}_{n}^{F}+\boldsymbol{\Delta}_{n}^{G}  \label{Non_eq3e}
\end{equation}
are the combined or the net non-equilibrium effects, where
\begin{equation}
\boldsymbol{\Delta}_{n}^{G} =\mathbf{M}_{n}(f_{ki})-\mathbf{M}
_{n}^{\text{eq}}(f_{ki}^{eq,\text{NEW}})
\end{equation}
are the non-equilibrium effects induced by the gradient force.
It is clear that, when the interparticle force disappears, the net non-equilibrium effects are only from the gradient force, i.e., $\boldsymbol{\Delta}_{n}=\boldsymbol{\Delta }_{n}^{G}$, corresponding to an ideal gas system.
Note that, $\mathbf{M}_{n}$ contain the information of $\mathbf{u}$, so do $\boldsymbol{\Delta}_{n}$ which describe both the HNE and TNE effects. If we use the central moment $\mathbf{M}_{n}^{*}(f_{ki})=\sum f_{ki}(\mathbf{v}_{ki}-\mathbf{u})^{n}$ which is only the representation of the thermo-fluctuations of molecules relative to $\mathbf{u}$,
then $\boldsymbol{\Delta}_{n}^{*}$ do not contain the effects of $\mathbf{u}$, describing only the TNE effects.

\begin{figure}[tbp]
\center {
\epsfig{file=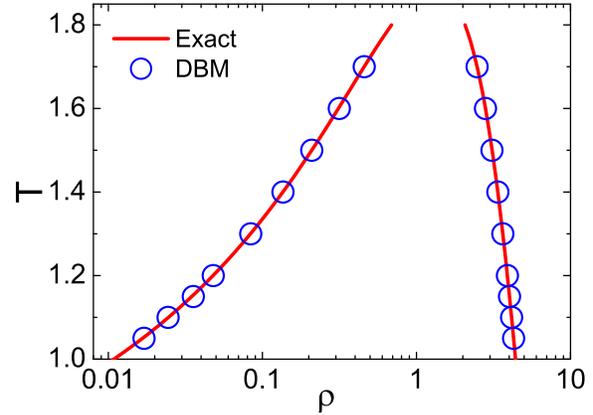,bbllx=0pt,bblly=0pt,bburx=581pt,bbury=420pt,
width=0.44\textwidth,clip=}}
\caption{Comparisons of the coexistence densities predicted by the DBM model and Maxwell constructions.}
\end{figure}
\begin{figure}[tbp]
\center {
\epsfig{file=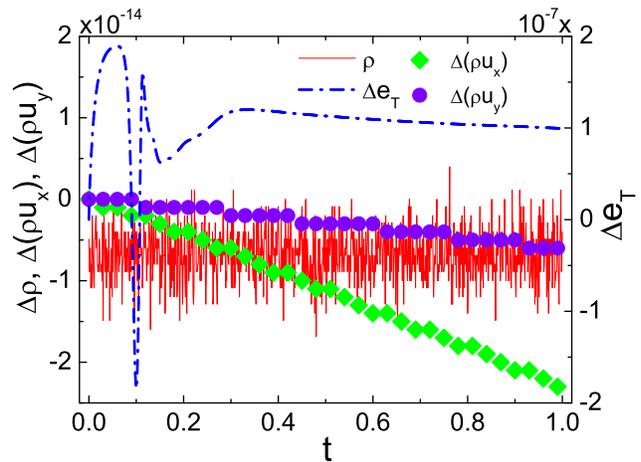,bbllx=3pt,bblly=2pt,bburx=581pt,bbury=424pt,
width=0.47\textwidth,clip=}}
\caption{Variations of $\rho$, $\rho \mathbf{u}$ and $e_{T}$ for a phase-separating process.}
\end{figure}
\begin{figure}[tbp]
\center {
\epsfig{file=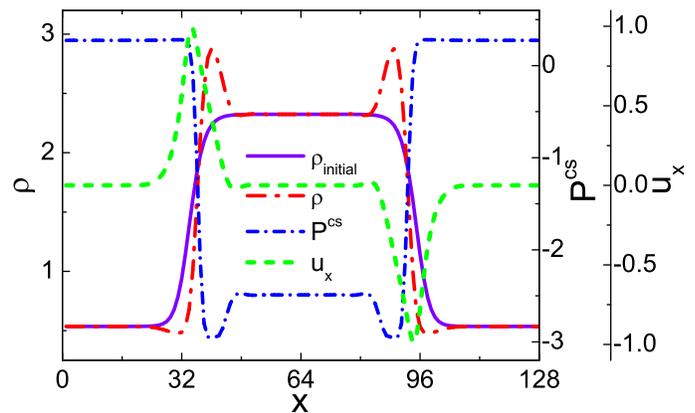,bbllx=11pt,bblly=6pt,bburx=574pt,bbury=351pt,
width=0.5\textwidth,clip=}}
\caption{Profiles of macroscopic quantities for the isothermal case at $t=0.01$.}
\end{figure}

\section{Simulation results and analysis}

In this section, we first validate the DBM model via two benchmarks; then investigate the HNE and TNE characteristics in both isothermal and thermal cases; finally, study the effects of surface tension on the thermal phase separation. Throughout our simulations, we set $a=2$ and $b=0.4$, then the critical density and temperature are $\rho ^{c}=1.30444$ and $T^{c}=1.88657$. The fast Fourier transform scheme with $16$-th order in precision\cite{Xu-PRE2011-EPL2012} and the second order Runge-Kutta scheme are utilized to discretize the spatial and temporal derivatives, respectively. The model parameters are $v_{1}=1.0$, $v_{2}=2.0$, $v_{3}=3.0$, $v_{4}=4.0$.

\subsection{Verification and validation}

To evaluate if the model can reproduce the correct thermodynamic equilibrium of the Carnahan-Starling system, we simulate the liquid-vapor coexistence curves at various temperatures with $128\times 1$ lattice and periodic boundary conditions in both directions. The initial conditions are $(\rho, T, \mathbf{u})=(\rho_{l}, 1.70, 0.0)$, if $N_{x}/4<x\leq3N_{x}/4$;
else $(\rho, T, \mathbf{u})=(\rho_{v}, 1.70, 0.0)$, where $\rho_{l}=2.473$ and $\rho_{v}=0.458$ are the theoretical liquid and vapor densities at $T=1
 .70$. Parameters are $\tau = 10^{-4}$, $\Delta x=\Delta y=2.222\times10^{-3}$, $\Delta t=2\times 10^{-5}$, $K=2.7\times10^{-5}$, $\zeta=0$, $q=-0.004$. The initial temperature is $1.70$ but drops by $0.01$ when the equilibrium state has been achieved. Figure 1 shows the phase diagram recovered from the DBM simulations and Maxwell constructions. The two sets of results are in accordance with each other, even when $T$ drops to $1.05$, corresponding to $R=\rho_{l}/\rho_{v}=255$. Clearly, it proves that the DBM is capable of handling multiphase flows with large density ratio as well as ensuring thermodynamic consistency.

Figure 2 illustrates variations of the density $\Delta \rho$, the momentum $\Delta (\rho \mathbf{u})$ and the total energy $\Delta e_{T}$ for a thermal phase-separating process calculated from the DBM model.
The initial conditions are $(\rho, T, \mathbf{u}) = (1.5+\Delta, 1.0, 0.0)$,
where $\Delta$ is a density noise with an amplitude of $0.001$. The remaining parameters are $N_{x}=N_{y}=128$, $\Delta x=\Delta y=5\times10^{-3}$, $\Delta t=5\times 10^{-5}$, $\tau=3\times10^{-3}$, $K=5\times10^{-5}$, $q=-0.002$. It is observed that, even when the initial state is quenched much lower than the critical temperature, $\Delta \rho$ and $\Delta (\rho \mathbf{u})$ maintain totally to machine accuracy. $\Delta e_{T}(t)$ fluctuates around its initial value when $t<0.4$, then keeps nearly to a constant. The initial fluctuation is due to the numerical discretization errors induced by the emergence of numerous interfaces during the spinodal decomposition stage. However, the maximum deviation of $e_{T}$ is $2\times10^{-7}$, indicating that the DBM is adequate to guarantee energy conservation.

\subsection{Non-equilibrium characteristics: isothermal and thermal cases}

\begin{figure*}[tbp]
\center {
\epsfig{file=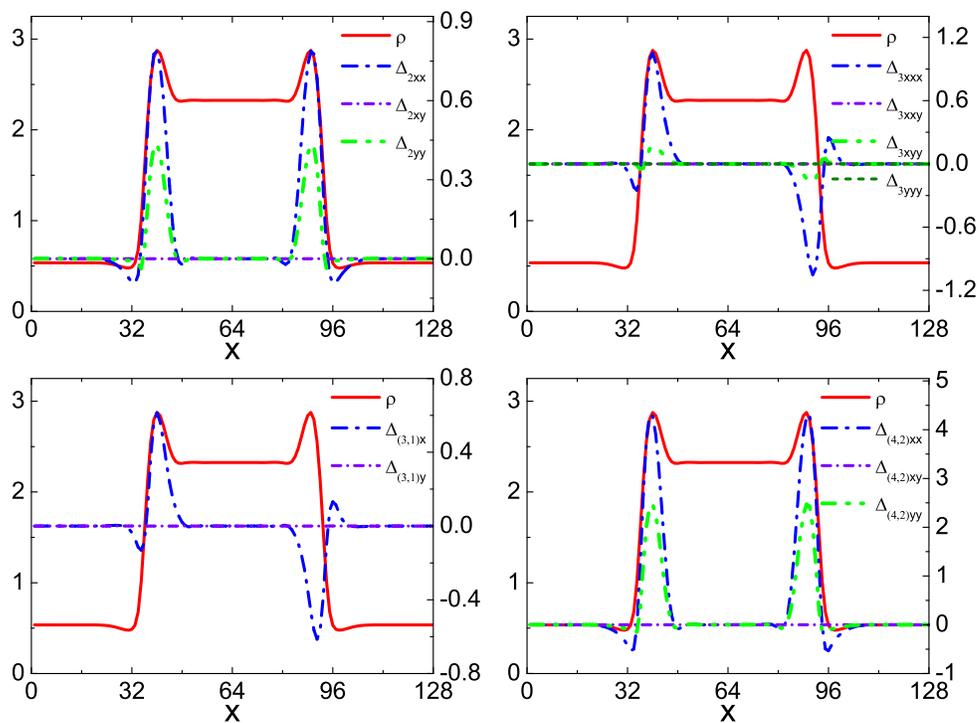,bbllx=109pt,bblly=284pt,bburx=568pt,bbury=624pt,
width=0.72\textwidth,clip=}}
\caption{The total non-equilibrium manifestations $\boldsymbol{\Delta}_{2}$, $\boldsymbol{\Delta}_{3}$, $\boldsymbol{\Delta}_{3,1}$ and $\boldsymbol{\Delta}_{4,2}$ for the isothermal case as shown in Fig. 3 (right Y-axis). The density profile at $t=0.01$ is also shown in each plot to guide the eyes (left Y-axis).}
\end{figure*}
\begin{figure*}[tbp]
\center {
\epsfig{file=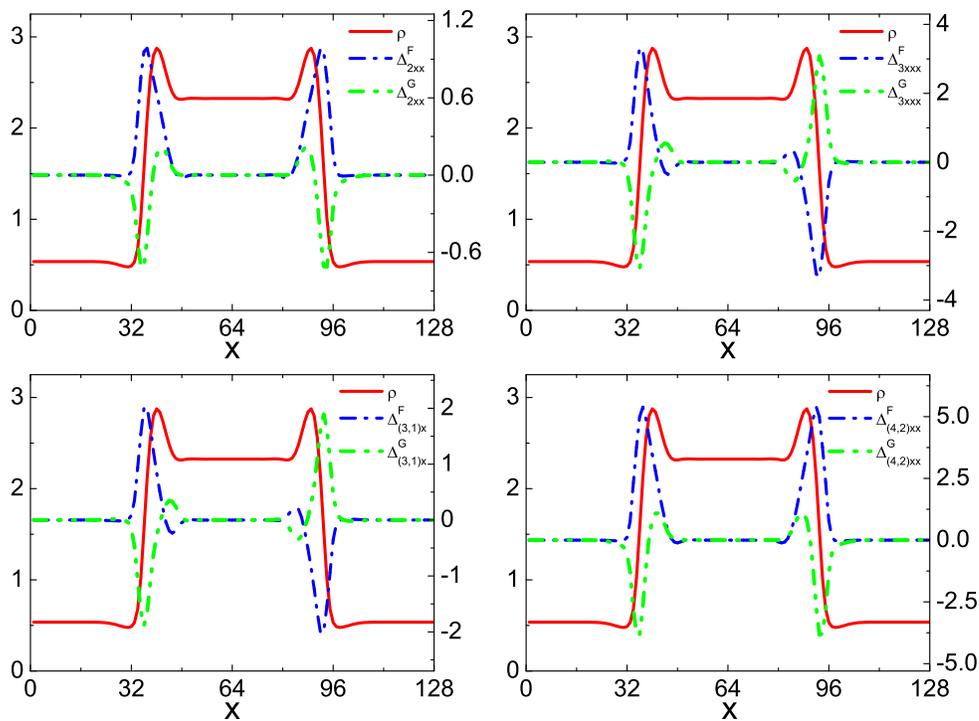,bbllx=109pt,bblly=282pt,bburx=568pt,bbury=624pt,
width=0.72\textwidth,clip=}}
\caption{The $x$ component of non-equilibrium manifestations induced by the interparticle force and the gradient force for the isothermal case at $t=0.01$ as shown in Fig. 3. The density profile at the same time is also shown to guide the eyes.}
\end{figure*}

For simplicity, we first examine the non-equilibrium effects in one-dimensional isothermal case. We set the equilibrium density profile at $T=1.74$ (see Fig. 3) as the initial state $\rho_{\text{initial}}(x)$.
When simulation starts, the system is suddenly quenched to $T=1.27$ and fixed at this temperature during the whole procedure. Physically, the isothermal results correspond to the case where the latent heat of phase transition approaches zero.
Parameters are consistent with what we used in Fig. 1 except for $\tau=10^{-3}$ and $K=1.1\times10^{-5}$. Profiles of the macroscopic quantities at $t=0.01$ are exhibited in Fig. 3.
We see the HNE as below. The decrease in temperature leads to the appearance of pressure gradients near the liquid-vapor interfaces which drive the vapor phase flows to the liquid side. As a result, the liquid (vapor) phase increases (decreases) its density, then the phase separation phenomenon takes place.

Figure 4 displays the total non-equilibrium manifestations $\boldsymbol{\Delta}_{2}$, $\boldsymbol{\Delta}_{3}$, $\boldsymbol{\Delta}_{3,1}$, and $\boldsymbol{\Delta}_{4,2}$ for Fig. 3, which suggests the following information during the procedure deviating from thermodynamic equilibrium:
(1) due to the initial fields are symmetric about the vertical line $x=N_{x}/2$, the non-equilibrium behaviors are also symmetric (for $\boldsymbol{\Delta}_{2}$ and $\boldsymbol{\Delta}_{4,2}$) or antisymmetric (for $\boldsymbol{\Delta}_{3}$ and $\boldsymbol{\Delta}_{3,1}$) about the same line;
(2) the non-equilibrium effects are mainly around the liquid-vapor interfaces where the gradients of macroscopic quantities and interparticle force arise, and attain their maxima (minima) at the point of the maximum density difference $\delta \rho_{\text{max}}$. This can be interpreted as follows.
In the first panel, according to the nature of $f^{eq}_{ki}$, we have $M_{2xx}^{eq}=\rho (T+u^{2}_{x})$, then $\Delta_{2xx}=M_{2xx}-M_{2xx}^{eq} \propto (T+u^{2}_{x})\delta \rho + \rho \delta(u^{2}_{x}) $. Quantitatively, $(T+u^{2}_{x})\delta \rho$ is the leading part of $\Delta_{2xx}$, then $\Delta_{2xx} \propto \delta \rho$ approximately. Therefore, when $\delta \rho>0$, $\Delta_{2xx}>0$, when $\delta \rho<0$, $\Delta_{2xx}<0$ (see the two troughs positioned at $x=32$ and $96$ for the vapor phases with decreasing densities). We also see that $\Delta_{2yy}<\Delta_{2xx}$. Numerically, it is because $\Delta_{2yy} \propto T\delta \rho$ (due to $u_{y}=0$), and physically, the non-equilibrium driving force $I_{ki}$ acts only along the $x$ axis and induces stronger non-equilibrium effects. Behaviors of $\boldsymbol{\Delta}_{4,2}$ can be analyzed in a similar way;
(3) $\Delta_{3xxx}$ shows a negative peak and a positive one with different amplitudes in the left half part of the computational domain, so do $\Delta_{3xyy}$ and $\Delta_{3,1x}$. Since $\Delta_{3xxx} \propto \delta (\rho u_{x})$, around the left interface $\delta u_{x}>0$ while $\delta \rho$ is negative at first, but positive later owing to phase separation, so at first $\Delta_{3xxx}<0$ and then $\Delta_{3xxx}>0$;
(4) there are some zero-components, such as $\Delta_{2xy}$, $\Delta_{3xxy}$, $\Delta_{3yyy}$, $\Delta_{3,1y}$, and $\Delta_{4,2xy}$. Physically, $\Delta_{2xy}$ accounts for the shear effects, $\Delta_{3xxy}+\Delta_{3yyy}=2\Delta_{3,1y}$ associate with the energy flux in the $y$ direction.
Since the system, or more fundamentally $f_{ki}(x,y)$, is symmetrical about the $y$ axis without gradients of macroscopic quantities, there are neither shear effects nor energy flux along this direction.

To probe the non-equilibrium effects induced by the interparticle force and the gradient force, respectively, we present the $x$ component of $\boldsymbol{\Delta}^{F}$ and $\boldsymbol{\Delta}^{G}$ in Fig. 5. Two remarkable features can be observed. Firstly, $\boldsymbol{\Delta}^{F}$ is opposite to and stronger than $\boldsymbol{\Delta}^{G}$. Physically, the interparticle force is the active force which drives the system evolution and increases the gradients of macroscopic quantities, while gradient force is a passive one dissipated partly by viscosity, thermal diffusion and surface tension, etc. The former is stronger than the latter before attainment of the final thermodynamic equilibrium state. Secondly, different from $\boldsymbol{\Delta}$, $\boldsymbol{\Delta}^{F}$ and $\boldsymbol{\Delta}^{G}$ achieve their maxima or minima at the middle of the interface, where the gradients of macroscopic quantities own the greatest values (except for $u_{x}$). This is correct since the forcing term is operating through gradients of macroscopic quantities.

To examine the effects of latent heat, now we go to the thermal case, shown in Fig. 6. Parameters and the initial state are consistent with what we used in the isothermal case. When simulation starts, the system is suddenly quenched to $T=1.0$, but $T(t)$ is free. So $\rho(x)$ and $u_{x}(x)$ exhibit similar behaviors with Fig. 3, except for $T(x)$. Due to the release (absorption) of latent heat, the temperature of the liquid (vapor) phase over the whole domain increases (decreases). This is the distinctive feature compared to isothermal case where latent heat is zero or exchanged with the connecting heat bath.
Figure 7 depicts the corresponding TNE quantity $\boldsymbol{\Delta}^{*}_{2}$ where the central moment $\mathbf{M}_{2}^{*}$ is employed. It is noteworthy that at $t=0.022$, the system is farthest away from the thermodynamic equilibrium and $|\boldsymbol{\Delta}^{*}_{2}|$ has the largest value. The trace of $\mathbf{M}_{2}^{*}$ associates with the internal kinetic energy. Compared to the first panel in Fig. 4, two prominent differences can be found. At first, in the thermal case $\Delta_{2xx}^{*}=-\Delta_{2yy}^{*}$, but in the isothermal case, $|\Delta_{2xx}^{*}|>|\Delta_{2yy}^{*}|$ (theoretically $\boldsymbol{\Delta}_{2}=\boldsymbol{\Delta}^{*}_{2}$). Comparisons of $\boldsymbol{\Delta}_{2}^{*}$ in both cases demonstrate that the internal kinetic energy are anisotropic in different degrees of freedom and the energy equipartition theory is broken down under the non-equilibrium case. While, in the thermal case, the internal kinetic energy at each point is conserved, thereby $\Delta_{2xx}^{*}+\Delta_{2yy}^{*}=0$.
Secondly, the non-equilibrium effects in isothermal case are much pronounced since the gradient force and gradients of macroscopic quantities are much stronger than those in the thermal case due to the loss of latent heat.

In Fig. 8, we exhibit the time evolutions of the averaged
$\bar{\Delta}_{2xx}^{* F}=\frac{1}{N_{x}N_{y}}\sum_{i,j}\Delta_{2xx}^{*F}(i,j)$ ($i=1,...,N_{x}; j=1,...,N_{y}$), the averaged
$\bar{\Delta}_{2xx}^{* G}$, and the averaged $\bar{\Delta}^{*}_{2xx}$ (defined similarly) for the thermal case, which reveals the following scenarios. Under the combined actions of the interparticle force and gradient force, the TNE behaviors appear, increase quickly and arrive at their maxima. The interparticle force is stronger than the gradient force, hence $|\boldsymbol{\Delta}^{*F}|>|\boldsymbol{\Delta}^{*G}|$.
These two kinds of driving forces influence, compete and balance partly with each other, resulting in the overshoot, oscillation and decay in $\boldsymbol{\Delta}^{*}$. Finally, when the system arrives at its steady state, i.e., the hydrodynamic equilibrium state, the two forces balance totally with each other, and consequently the net TNE effects vanish, $\boldsymbol{\Delta}^{*}=\boldsymbol{\Delta}^{*G}+\boldsymbol{\Delta}^{*F}=0$.

\begin{figure}[tbp]
\center {
\epsfig{file=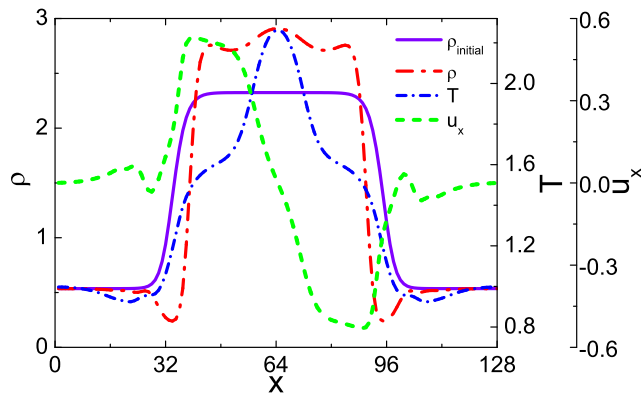,bbllx=12pt,bblly=4pt,bburx=574pt,bbury=350pt,
width=0.47\textwidth,clip=}}
\caption{Profiles of macroscopic quantities for the thermal case at $t=0.022$.}
\end{figure}
\begin{figure}[tbp]
\center {
\epsfig{file=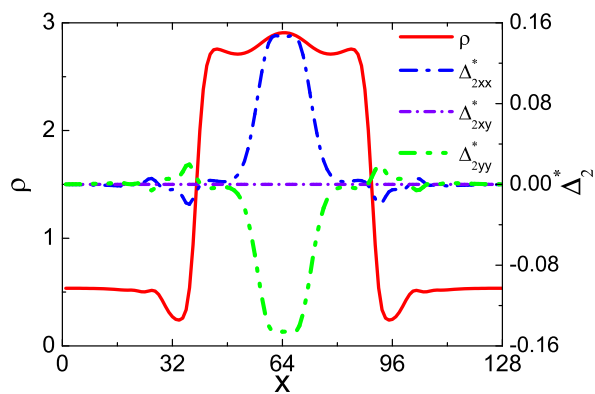,bbllx=7pt,bblly=2pt,bburx=574pt,bbury=380pt,
width=0.44\textwidth,clip=}}
\caption{TNE manifestations $\boldsymbol{\Delta}^{*}_{2}$ for the thermal case at $t=0.022$. It is noteworthy that, at this moment, the system is farthest away from the thermodynamic equilibrium and $|\boldsymbol{\Delta}^{*}_{2}|$ has the largest value.
The density profile at the same time is also shown to guide the eyes.}
\end{figure}
\begin{figure}[tbp]
\center {
\epsfig{file=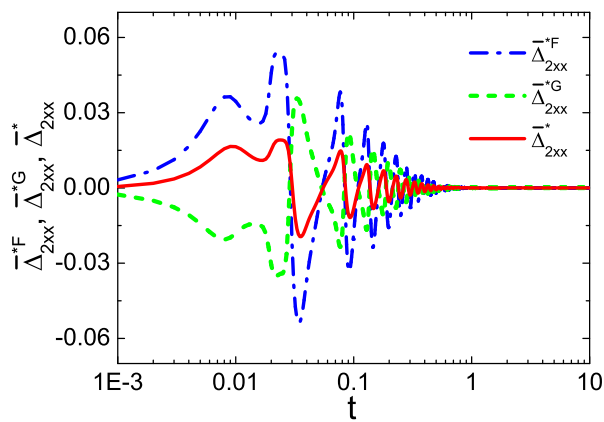,bbllx=0pt,bblly=0pt,bburx=565pt,bbury=412pt,
width=0.44\textwidth,clip=}}
\caption{Time evolutions of $\bar{\Delta}^{*F}_{2xx}$, $\bar{\Delta}^{*G}_{2xx}$ and $\bar{\Delta}^{*}_{2xx}$ for the thermal case.}
\end{figure}

\subsection{Effects of surface tension on thermal phase separation}

\begin{figure*}[tbp]
\center {
\epsfig{file=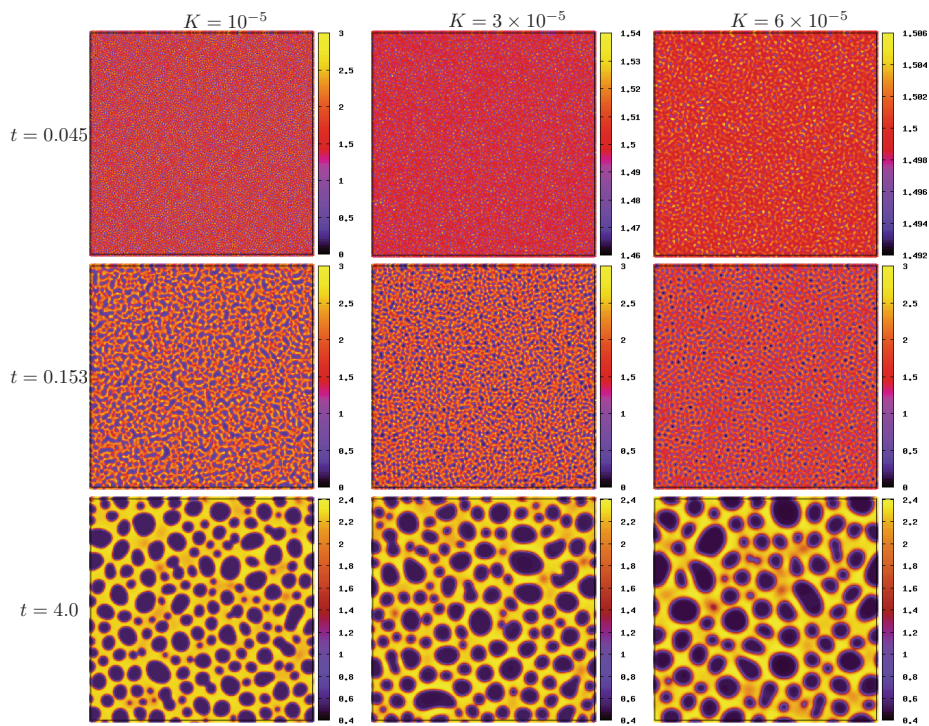,bbllx=0pt,bblly=0pt,bburx=483pt,bbury=374pt,
width=0.7\textwidth,clip=}}
\caption{Density patterns at three representative times during thermal phase separation processes, $t=0.045$ (the first row), $t=0.153$
(the second row) and $t=4.0$ (the third row). From left to right, the three columns
correspond to cases with $K=10^{-5}$, $3\times 10^{-5}$ and $6\times10^{-5}$, respectively.}
\end{figure*}

\begin{figure*}[tbp]
\center {
\epsfig{file=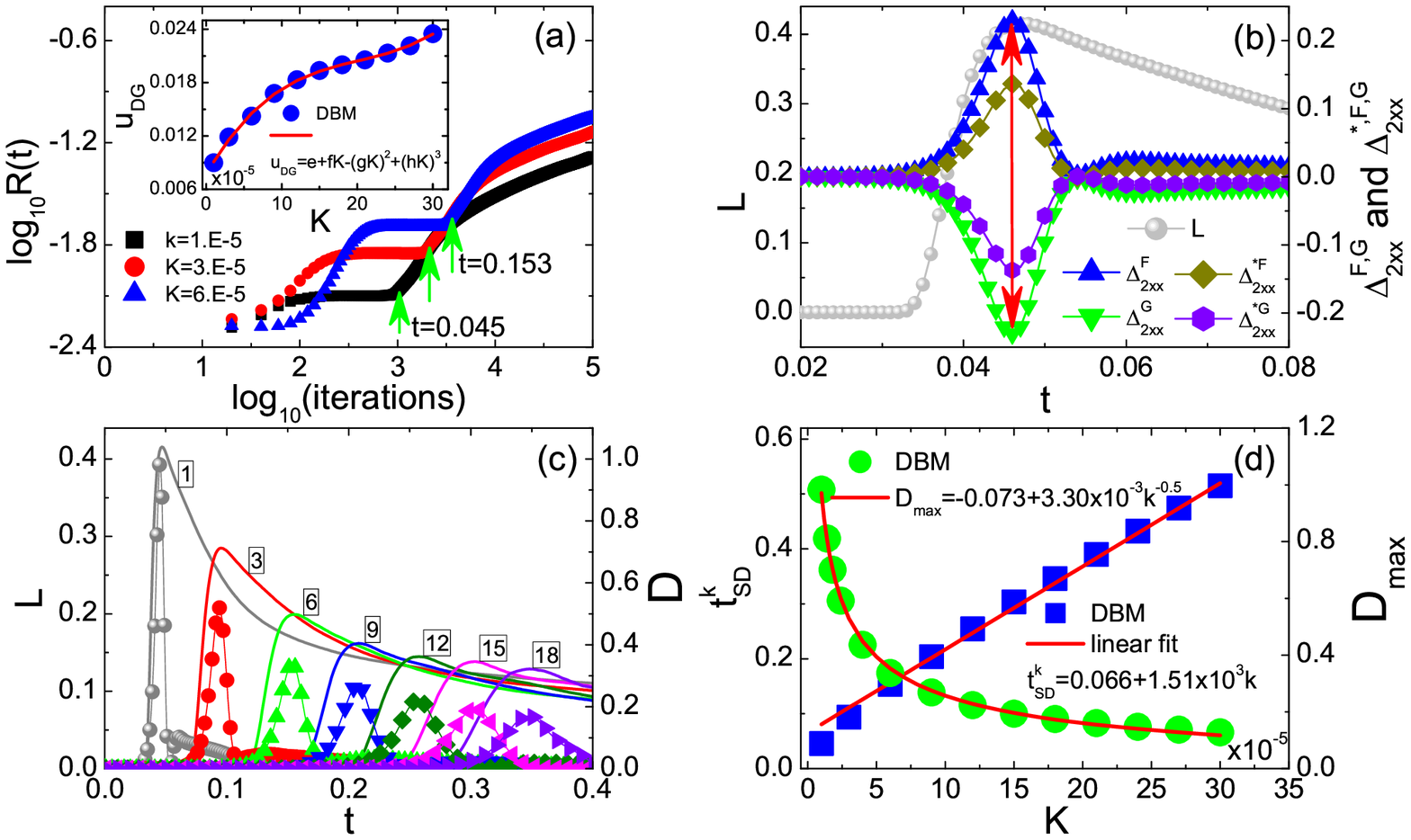,bbllx=0pt,bblly=0pt,bburx=574pt,bbury=348pt,
width=0.99\textwidth,clip=}}
\caption{(a) Evolutions of the characteristic domain sizes $R$ for the procedures shown in Fig. 9.
(b) Evolutions of the boundary length $L$ and the $xx$ component of some TNE manifestations for the phase separation process with $K=10^{-5}$. (c) Evolutions of the boundary lengths $L$ (solid curves) and the corresponding TNE intensities $D$ (curves with solid symbols) for phase separation processes with various $K$. Here $1,3,6,...,18$ labeled on the $L$-curves indicate cases with $K=10^{-5}$, $3\times10^{-5}$, $6\times 10^{-5}$,...,$1.8 \times 10^{-4}$, respectively.
(d) Duration of the spinodal decomposition stage $t_{SD}$ and the maximum TNE intensity $D_{\text{max}}$ as functions of $K$.}
\end{figure*}
It is well known that, when a system is instantaneously quenched from a disordered state into a coexistence one,
the fluids undergo two TNE stages:\cite{Succi,Xu-PRE2011-EPL2012,EPJB-Sofonea} the early spinodal decomposition stage and the later domain growth stage, then approach the finial totally separated equilibrium state.
Previous studies focus mainly on the domain growth law at the second stage.\cite{Succi,Gonnella-group} Due to the emergence of large variety of complex spatial patterns during phase separation, especially during the spinodal decomposition stage, how to exactly distinguish
the two stages is an open problem; aside from, the TNE behaviors during the whole process are barely concerned. Recently, with the help of Minkowski measures, we presented a numerical criterion for exactly separating the two stages and quantitatively investigated the heat conduction, viscosity
and Prandtl number effects on thermal phase separation.\cite{Xu-PRE2011-EPL2012} Here we further give out a physical criterion and investigate the effect of surface tension.

To that aim, we run simulations on symmetric phase separations with various surface tension coefficients $K$ on periodic $512\times 512$ domains. The initial conditions and parameters are consistent with those used in Fig. 2.
Figure 9 shows the density patterns with various $K$ at three representative times. From left to right, the three columns
correspond to cases with $K=10^{-5}$, $3\times 10^{-5}$ and $6\times10^{-5}$, respectively. Figure 9 manifests that surface tension strongly affects the pattern morphology, the speed and the depth of phase separation procedure. More precisely, at $t=0.045$, for the case with small $K=10^{-5}$, numerous mini domains with large density difference, separated by complicated interfaces, appear, suggesting that the procedure has already entered the final spinodal decomposition stage. While for cases of larger $K$, the density variance is quite small, decreasing with increasing $K$. Nevertheless when $t=0.153$, the averaged domain size and the phase separation depth for the three cases are almost the same; all cases proceed to the domain growth stage. As time evolves further, we observe that the larger the $K$, the faster the phase separation, the bigger the averaged size, the fewer the number of domains and the wider the interface. Summarizing, Fig. 9 demonstrates that the surface tension effects prolong the spinodal decomposition stage but
accelerate the domain growth stage.

These results are further confirmed by the time history of the characteristic domain size $R(t)$,\cite{Succi,Gonnella-group,Xu-PRE2011-EPL2012} plotted in Fig. 10(a). The $R(t)$ curves behave similarly and distinguish approximately the phase separation process into two stages. At the first stage, it increases and arrives at a platform marked by the green arrow. In fact, the marked point corresponds to the end of the spinodal decomposition stage. Vladimirova \emph{et al.}\cite{Vladimirova} pointed out that the plateau depends on the depth of temperature quench and the intensity of random noise. Here we find, it also depends on the surface tension. The larger the surface tension, the longer the duration $t_{SD}$ of the spinodal decomposition stage, and the larger the domain size for the spinodal decomposition stage $R_{SD}$. Our results are consistent with the theoretical analysis, neglecting heat conduction and viscous effects.\cite{K-SD}
Essentially, phase separation is a process, through which the potential energy transforms into the thermal energy and the interfacial energy. Under the action of interparticle force, a liquid (vapor) embryo is continuously gaining (losing) molecules due to condensation (evaporation), then the interface emerges and part of the potential energy transforms into the interfacial energy that is proportional to $K$. Therefore, an increasing $K$ means an increasing interfacial energy, an increasing $t_{SD}$ required for completing such an energy conversion process. On the other hand, the surface tension always resists the appearance of new interface to minimize the interfacial energy. The larger the surface tension is, the stronger the resistance is and the longer it takes for sharp interfaces to form.

Afterwards, in the domain growth stage, under the action of surface tension, small domains merge together to minimize the free energy, which naturally leads to the continuous growth in $R(t)$. The slopes of $R(t)$ curves, corresponding to the phase separation speeds during the domain growth stage $u_{DG}$, increase with $K$.
Therefore, at the domain growth stage, the phase separation process is remarkably accelerated by the surface tension. Specifically, the $R(t)$ curve with $K=6\times10^{-5}$ crosses with the other two at $t=0.153$, then rises quickly and exceeds the former two.
When $K$ varies from $10^{-5}$ to $3\times10^{-4}$, the dependence of $u_{SD}$ on $K$ can be fitted by
\begin{equation}
u_{DG}=e+fK-(gK)^{2}+(hK)^{3},
\end{equation}
with $e=0.00764$, $f=1.51\times 10^{2}$, $g=8.06\times 10^{2}$, $h=1.02\times 10^{3}$, as shown in the legend of Fig. 10(a).
Our results show qualitative agreement with theoretical analysis,\cite{JFM} simulations by smoothed particle hydrodynamics,\cite{SPH} and lattice Boltzmann study\cite{LB-KK} for isothermal case.

To accurately determine the $t_{SD}$, in Fig. 10(b) we monitor the time evolution of the second Minkowski measure: boundary length $L(t)$ for the density threshold $\rho_{th}=1.70$ for which the density pattern has the largest boundary length. Meanwhile, some TNE manifestations are exhibited in the same panel. It is interesting to note that the peak of the $L(t)$ curve exactly coincides with the peaks or troughs of the TNE curves. Each nonzero component of $\boldsymbol{\Delta}$ or $\boldsymbol{\Delta}^{*}$ describes the TNE from its own side. To roughly and averagely estimate the deviation amplitude from the thermodynamic equilibrium, we further define a ``TNE strength"
\begin{equation}
D=\sqrt{\boldsymbol{\Delta_{2}^{*2}}+\boldsymbol{\Delta_{3}^{*2}}+\boldsymbol{\Delta_{3,1}^{*2}}+\boldsymbol{\Delta_{4,2}^{*2}}}.
\end{equation}
We may also use $\sqrt{\boldsymbol{\Delta_{2}^{2}}+\boldsymbol{\Delta_{3}^{2}}+\boldsymbol{\Delta_{3,1}^{2}}+\boldsymbol{\Delta_{4,2}^{2}}}$ (or its $F$, or $G$ component). In general, the DBM equation is dimensionless, so do $\boldsymbol{\Delta}$ and $D$. $D=0$ indicates that the system is in thermodynamic equilibrium and $D>0$ out of the thermodynamic equilibrium. Shown in Fig. 10(c) are the time evolutions of $L(t)$ (solid curves) and $D(t)$ (curves with solid symbols, calculated from $\boldsymbol{\Delta}^{*F}$) for various $K$, where $1,3,6,...,18$ labeled on the $L(t)$-curves indicate cases with $K=10^{-5}$, $3\times10^{-5}$, $6\times 10^{-5}$,...,$1.8 \times 10^{-4}$, respectively.
To be seen is a perfect coincidence between the peaks of $L(t)$ and $D(t)$ in pairs. Therefore, the time evolution of $D(t)$ provides a convenient, efficient and physical way to divide reasonably the spinodal decomposition and the domain growth stages. The left (right) part of the peak corresponds to the spinodal decomposition (domain growth) stage. In addition, compared to the morphological way, the extension of the current approach to three dimensions is straightforward.

Figure 10(c), again, demonstrates our conclusions: during the spinodal decomposition stage, the larger the surface tension, the longer the time delay, the smaller the slope of the TNE curve, and the weaker the TNE intensity. For the case with larger $K$, the longer time delay and the subsequent relatively mild increase in $D$ makes the $t_{SD}$ longer. For example,
when $K=10^{-5}$, $t_{SD}=0.045$, while when $K$ increases to $1.8\times10^{-4}$, $t_{SD}$ increases significantly to $0.35$. When $K$ varies in the range $[10^{-5}, 3\times10^{-4}]$, the dependence of $t_{SD}$ on $K$ can be fitted with the following form
\begin{equation}
t_{SD}=a+bK\text{,}  \label{t_SD-k}
\end{equation}
with $a=0.066$ and $b=1.51\times 10^{3}$, as shown in Fig. 10(d).
Furthermore, due to the interface, the phase separation depth, as well as the gradient force and interparticle force achieve their peak values at the end of the spinodal decomposition stage, the TNE intensity is the strongest at this moment.
Nevertheless, it is also found that the surface tension effects decrease the maximum of the TNE strength $D_{\text{max}}$ approximately in the following way
\begin{equation}
D_{\text{max}}=c+dK^{-0.5}\text{,}  \label{dmax-k}
\end{equation}
with $c=-0.073$ and $d=3.30\times 10^{-3}$, as shown in Fig. 10(d).
Physically, the Knudsen number is usually employed to classify the level of TNE, which is defined as the ratio between the molecular mean-free-path $\lambda$ and a character length $L$ at which macroscopic variations are of interest. For a phase separation process, we can take $L$ to be roughly the domain size at the end of the spinodal decomposition stage, $R_{SD}$. \cite{RSD} Thus the mean Knudsen number $\text{Kn}=\lambda/2R_{SD}$.
As displayed in Fig. 10(a), $R_{SD}$ increases with $K$, thus, $\text{Kn}$ and the TNE strength decrease with $K$ oppositely. Numerically, a larger $K$ will broaden the interface width, reduce the gradient force and refrain the TNE intensity.

\section{Conclusion}
An energy-conserving discrete Boltzmann model for multiphase flow system with flexible density ratio is developed and utilized to study both the hydrodynamic non-equilibrium and thermodynamic non-equilibrium effects in phase separation processes. Efficient parallel implementation and ability to capture the non-equilibrium effects are the two advantages of the discrete Boltzmann model on computational and physical sides, respectively. Besides being helpful for better understanding the hydrodynamic non-equilibrium behaviors in the phase separation process, the thermodynamic non-equilibrium effects permit to formulate a physical criterion to separately analyze the spinodal decomposition and domain growth stages.
This work marks a preliminary step towards a deeper understanding of
hydrodynamic and thermodynamic non-equilibrium effects on
phase-separation phenomena. Much scope is left for future
investigations in the field.

\section*{Acknowledgments}
We are grateful to the anonymous referees for their valuable comments and suggestions.
We warmly thank Drs. Huilin Lai and Chuandong Lin for many helpful discussions.
We acknowledge support of the Science Foundation of National Key Laboratory of Computational Physics, the Open Project Program of State Key Laboratory of Theoretical Physics, Institute of Theoretical Physics, Chinese Academy of Sciences (Y4KF151CJ1), National Natural Science Foundation of China (11475028,11202003 and 11203001), Science Foundation of Hebei Province
(A2013409003, YQ2013013 and ZD2014089).

\balance
\footnotesize{
\providecommand*{\mcitethebibliography}{\thebibliography}
\csname @ifundefined\endcsname{endmcitethebibliography}
{\let\endmcitethebibliography\endthebibliography}{}

}

\end{document}